\documentclass{llncs}

\usepackage{booktabs}
\usepackage{subcaption}
\usepackage{amssymb}
\usepackage{amsmath}
\usepackage[dvipsnames]{xcolor}
\usepackage{multirow}
\usepackage[linesnumbered]{algorithm2e}
\usepackage{multicol}
\usepackage{mdframed}
\usepackage{listings}
\usepackage{hyperref}
\usepackage{tikz}

\usetikzlibrary{shapes.geometric, arrows}

\definecolor[named]{lipicsLightGray}{rgb}{0.85,0.85,0.86}

\newcommand{\tool}[1]{\textsf{#1}}
\newcommand{\aligator}{{\tool{Aligator}}\xspace}
\newcommand{\absynth}{{\tool{Absynth}}\xspace}

\newcommand{\pp}{\pi}

\newcommand{\tup}[1]{\langle #1 \rangle}

\newcommand{\id}{\tup}

\newcommand{\reserved}[1]{\textbf{\underline{#1}}} 

\newcommand{\DO}{\reserved{do}}

\newcommand{\WHILE}{\reserved{while}}

\newcommand{\END}{\reserved{end}}

\newcommand{\N}{\mathbb{N}}

\renewcommand{\vec}[1]{\boldsymbol{#1}}

\begin{document}

\mainmatter
\pagestyle{headings}

\title{Algebra-based Synthesis of Loops and their Invariants (Invited Paper)}
\titlerunning{Algebra-based Synthesis of Loops and their Invariants}

\author{Andreas Humenberger \and Laura Kov\'acs}

\authorrunning{L.~Kov\'acs}

\institute{
TU Wien}

\maketitle

\begin{abstract}
Provably correct software is one of the key challenges in our software-driven society.
While formal verification
establishes the correctness of a given program, the result of program synthesis
is a program which is correct by construction.
In this paper we overview  some of our results for both of these scenarios when
analysing programs with loops. 
The class of loops we consider can be modelled by a system of
linear recurrence equations with constant coefficients, called
C-finite recurrences. We first describe an algorithmic approach for
synthesising all polynomial equality invariants of such non-deterministic numeric single-path
loops. By reverse engineering invariant synthesis, we then
describe an automated method for synthesising program loops satisfying
a given set of polynomial loop invariants. Our results have  applications
towards proving partial correctness of programs, compiler optimisation
and generating number sequences from algebraic relations.

\end{abstract}


\section{Introduction}

 The two most
rigorous approaches for providing correct software are given by formal program
verification and program synthesis~\cite{GulwaniPOPL10}. The task of formal
verification is to prove correctness of a given program with respect to a given
logical specification~\cite{Hoare69,CousotC77,Clarke81}. 
On the other hand, program synthesis aims at generating programs which adhere to
a given specification~\cite{MannaW80,Alur15}. The result of a synthesis problem is therefore a program
which is correct by construction with respect to the specification. While formal verification has received
considerable attention with impressive results, for example, in ensuring safety
of device drivers~\cite{SLAM11} and security of web services~\cite{Cook18},
program synthesis turns out to be an algorithmically much more difficult
challenge~\cite{KuncakACM12}.

Both in the setting of verification and synthesis, one of the main
challenges is to verify or synthesise programs with
loops/recursion. In formal verification,
solving this challenge requires for example \emph{synthesising loop invariants}
~\cite{Rodriguez-CarbonellK07,HumenbergerJK17,KincaidCBR18}.
Intuitively, a loop invariant is a formal description of the behaviour of the
loop, expressing loop properties that hold at arbitrary loop
iterations. For the purpose of automating formal verification,
synthesising loop invariants that are
inductive is of critical importance, as inductive invariants describe
program properties/safety assertions that hold
before and after each loop iteration. 

In program synthesis, reasoning with loops requires answering the
question whether there exists a loop satisfying a given loop invariant and
synthesising a loop with respect to a given invariant. We refer
to this task of synthesis as {\it loop synthesis}. As such, we
consider loop synthesis as the reverse problem of loop invariant generation/synthesis: rather than
generating invariants summarising a given loop, we synthesise loops whose
summaries are captured by a given invariant property.

In this paper, we overview algebra-based algorithms for
automating reasoning about loops and their invariants. The key
ingredients of our work come with deriving and solving
algebraic recurrences capturing the functional behaviour of loops to
be verified and/or synthesised. To this end, 
we consider additional
requirements on the loops to be verified/synthesised, in particular by imposing syntactic
constraints on the form of loop expressions. The imposed
constraints allow us to reduce the verification/synthesis task to the
problem of solving algebraic recurrences of special forms. Here, we mainly focus on loops whose functional summaries are precisely
captured by so-called 
C-finite 
recurrences~\cite{KauersP11}, that is linear recurrences with
constant coefficients, for which closed form solutions always exist.
We use symbolic summation techniques over C-finite 
recurrences to compute closed forms and combine these closed forms
with additional constraints to ensure  that (i) algebraic
relations among closed forms yield polynomial loop invariants and (ii)
loops synthesised from such polynomial loop invariants implement only
affine assignments.

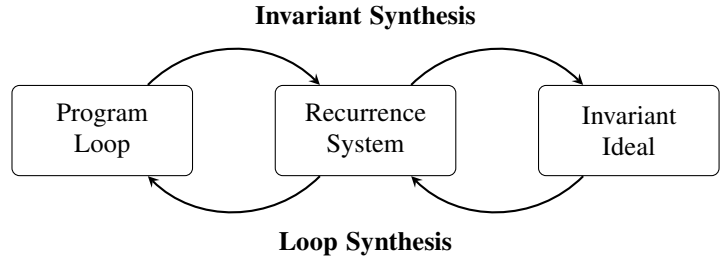
\begin{figure}[t]
    \centering
    \tikzstyle{defnode} = [rectangle, rounded corners=1mm, inner xsep=2mm, text width=2.0cm, minimum width=1.7cm, minimum height=1.2cm, text centered, draw=black, xshift=2.0cm, font=\normalsize]
    \tikzstyle{desc} = [rectangle, rounded corners=1mm, inner xsep=2mm, text width=4.6cm, minimum width=1.7cm, minimum height=1cm, text centered, font=\normalsize\bfseries]

    \begin{tikzpicture}[node distance=1.5cm]
        \node (loop)   [defnode]                  {Program Loop};
        \node (recsys) [defnode, right of=loop]   {Recurrence System};
        \node (ideal)  [defnode, right of=recsys] {Invariant Ideal};

        \node (invgen) [desc, above of=recsys] {Invariant Synthesis};
        \node (synthesis) [desc, below of=recsys] {Loop Synthesis};

        \path[thick,->,>=stealth]
            (loop)   edge[bend right=-45] node [right] {} (recsys)
            (recsys) edge[bend right=-45] node [right] {} (ideal);
        \path[thick,->,>=stealth]
            (ideal)  edge[bend right=-45] node [right] {} (recsys)
            (recsys) edge[bend right=-45] node [right] {} (loop);

    \end{tikzpicture}
    \caption{Algebra-based synthesis of loops and their invariants\label{fig:overview}.}
\end{figure}

Figure~\ref{fig:overview} overviews our
approach towards synthesising loops and/or their invariants.
In order to generate invariants, we extract a system of C-finite recurrence
equations describing loop updates. 
We then compute the polynomial ideal, called the {\it polynomial invariant
ideal},  containing all
polynomial equality invariants of the loop, by using recurrences
solving and Gr\"obner basis computation~\cite{Buchberger06}.
Any polynomial invariant of the given loop is then a logical
consequence of the polynomials from the computed polynomial ideal basis~\cite{kovacs08}. 
On the other hand, for loop synthesis, we take a basis of
the polynomial invariant ideal generated by given polynomial loop invariants  and
construct a polynomial constraint problem. This constraint problem precisely
characterises the set of all C-finite recurrence systems for which the
given polynomial invariants yield algebraic relations among the
induced 
C-finite number sequences. Every
solution of the constraint problem gives thus rise to a system of
C-finite recurrence equations,  which
is then turned into a loop for which the given polynomial relations
are 
loop invariants~\cite{Humenberger20}.

In the rest of this paper, we first motivate our results on examples for invariant and loop synthesis
(Section~\ref{sec:motivating}). We then 
report on 
algebra-based approaches for invariant generation
(Section~\ref{sec:InvGen}) and loop synthesis
(Section~\ref{sec:LoopSynt}), by summarising our main 
results published at~\cite{kovacs08,Humenberger20}.

\section{Motivating Examples for Synthesising Invariants and Loops}\label{sec:motivating}

\begin{figure}[t]
    \captionsetup[sub]{skip=2pt}
    \begin{subfigure}{.49\textwidth}
        \[
          \begin{tabular}{l}
            \reserved{requires} $N>0$\\[.5em]
 $(x, y, z) \gets (0, {0, 0})$\\
      $\WHILE~y<N~\DO$\\
        \quad$x \gets x+\colorbox{white}{$z+1$}$\\
        \quad $z \gets z+ 2$\\
      \quad$y \gets y+1$\\
      $\END$\\[.5em]
        \reserved{ensures} $x=N^2$
        {\color{white} aaa}
      \end{tabular}
        \]
        \caption{Invariant synthesis for partial correctness.}
        \label{fig:motivating:a}
    \end{subfigure}
    \begin{subfigure}{.49\textwidth}
        \[
          \begin{tabular}{l}
                        \reserved{requires} $N>0$\\[.5em]
      $(x, y) \gets (0, {0})$\\
      $\WHILE~y<N~\DO$\\
      \quad$x \gets x+ \colorbox{white}{$2y +1$}$\\
      \quad$y \gets y+1$\\
           $\END$\\\newline\\[.5em]
        \reserved{ensures} $x=N^2$
      \end{tabular}
        \]
        \caption{Loop synthesis~\label{fig:motivating:b} to ``optimize''
    Figure~\ref{fig:motivating:a}.}
    \end{subfigure}
    \caption{Motivating example for invariant and loop synthesis\label{fig:motivating}.}
\end{figure}

\noindent\paragraph{Loop Invariant Synthesis.}
Verifying safety conditions and establishing partial correctness of programs is
one use case of invariant generation. Consider for example the
program in Figure~\ref{fig:motivating:a}, annotated with pre- and
post-conditions specified respectively by the {\tt requires} and {\tt ensures} 
constructs. 
The program of Figure~\ref{fig:motivating:a} is clearly safe
as the post-condition is satisfied when the loop is exited.
However, to prove program safety 
we need additional
loop properties, i.e. inductive loop invariants,
that hold at any loop iteration. It is not hard to derive that after any
iteration $n$ of the loop (assuming $0\leq n \leq N$), the linear
invariant relation $y\leq N$ holds. It is also not
hard to argue that, upon exiting the loop, the value of $y$ is $N$.
However, such properties do not give us much
information about the (integer-valued) program variable $x$.
For proving program safety, we need to derive loop invariants relating
the values of $x, y, z$ at an arbitrary loop iteration $n$. 
Our work in~\cite{kovacs08}  generates 
such loop invariants by computing the polynomial ideal
${I=\id{x-y^2, z-2y}}$ as the so-called {\it polynomial
  invariant ideal}. The conjunction $x=y^2
\ \wedge z=2y$ of the polynomial relations  corresponding to the basis polynomials of $I$ is an
inductive loop invariant, which together with the invariant $y\leq N$
is sufficient to prove partial correctness of
Figure~\ref{fig:motivating:a}.

\begin{figure}[tb]
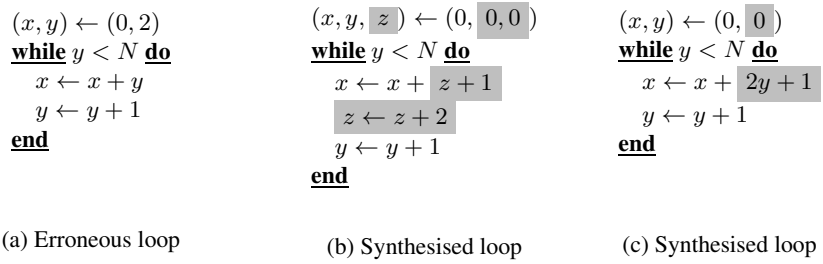

  \begin{subfigure}{.33\textwidth}
    \begin{center}
      \begin{tabular}{l}
      $(x, y) \gets (0, 2)$\\
      $\WHILE~y<N~\DO$\\
        \quad$x \gets x+y$\\
      \quad$y \gets y+1$\\
        $\END$\\
        {\color{white} aaa}
      \end{tabular}
    \end{center}
  \caption{Erroneous loop}\label{fig:Dafny:a}
  \end{subfigure}\hfill
 \begin{subfigure}{.33\textwidth}
    \begin{center}
      \begin{tabular}{l}
      $(x, y, \colorbox{lightgray}{$z$}) \gets (0, \colorbox{lightgray}{$0, 0$})$\\
      $\WHILE~y<N~\DO$\\
        \quad$x \gets x+ \colorbox{lightgray}{$z+1$}$\\
        \quad \colorbox{lightgray}{$z \gets z+ 2$}\\
      \quad$y \gets y+1$\\
      $\END$
      \end{tabular}
    \end{center}\caption{Synthesised loop}\label{fig:Dafny:c}
  \end{subfigure}
  \begin{subfigure}{.3\textwidth}
    \begin{center}
      \begin{tabular}{l}
      $(x, y) \gets (0, \colorbox{lightgray}{$0$})$\\
      $\WHILE~y<N~\DO$\\
      \quad$x \gets x+ \colorbox{lightgray}{$2y +1$}$\\
      \quad$y \gets y+1$\\
        $\END$\\
        {\color{white} aaa}
      \end{tabular}
    \end{center}
  \caption{Synthesised loop}\label{fig:Dafny:b}
  \end{subfigure}\hfill
  \caption{Program repair via loop
    synthesis. 
    Figures~\ref{fig:Dafny:c}-\ref{fig:Dafny:b}, corresponding also to the
    programs of Figures~\ref{fig:motivating:a}-\ref{fig:motivating:b}, are
    revised versions of Figure~\ref{fig:Dafny:a} such that
    ${x=y^2}$ is an invariant of
    Figures~\ref{fig:Dafny:c}-\ref{fig:Dafny:b}.}
  \label{fig:Dafny}
\end{figure}

\noindent\paragraph{Loop Synthesis.} 
One use case of loop synthesis is program optimisation. To reduce execution time
spent within loops, compiler optimisation techniques, such as strength
reduction~\cite{SRed01}, aim at replacing expensive loop operations with semantically
equivalent but less expensive operations and/or reducing the number of
 loop variables used within
loops. The burden of program optimisation in the presence of loops comes
however with identifying inductive loop variables and invariants to be used for
loop optimisation. Coming back to the loop in
Figure~\ref{fig:motivating:a}, as argued before, $x=y^2 \wedge \ z=2y
\wedge y\leq N$ is a loop
invariant of Figure~\ref{fig:motivating:a}. Moreover, only $x=y^2$ is
already a 
loop invariant of Figure~\ref{fig:motivating:a}.
Our loop synthesis procedure can be used to synthesise the affine 
loop of Figure~\ref{fig:motivating:b} from the polynomial invariant $x=y^2$, such that the synthesised
loop uses less variables and arithmetic operations than
Figure~\ref{fig:motivating:a}.
Note that program repair can also be considered as an instance of
program optimisation: while maintaining a given polynomial loop invariant, the
task is to revise and repair a given program such that it satisfies the
given invariant. Our synthesis approach therefore also provides a solution
to program repair, as illustrated in Figure~\ref{fig:Dafny}.

\section{Algebra-based Synthesis of Loop Invariants}\label{sec:InvGen}

\paragraph{Overview of state-of-the-art.}
One of the most related approaches to our work in automating the
synthesis of polynomial loop invariants comes with the seminal work of~\cite{GermanW75},
where a 
method for refining a user-given partial invariant was introduced to
prove partial correctness of a given program. 
One of the first fully automatic invariant generation
procedures was then given by~\cite{Karr76} for inferring affine relations within
affine programs. Since then, loop invariant generation was intensively
studied and the level of automation and expressivity with
respect to programs and their invariants steadily increased. Here we
overview the most related techniques to our work.

The approach of~\cite{Muller-OlmS04} generalised~\cite{Karr76}
and provided a method for computing all polynomial equality relations for affine
programs up to an a priori fixed degree. Recently, \cite{WorrellJACM}
constructively proved that the set of all polynomial equality invariants is
computable for affine programs.

The works of~\cite{SankaranarayananSM04} and~\cite{OliveiraBP16} fix a
polynomial template invariant and derive a constraint problem that encodes
properties of loop invariants, such as inductiveness. These constraint problems
are then solved by linear or polynomial algebra.
The methods of~\cite{ipl/Muller-OlmS04} and~\cite{scp/Rodriguez-CarbonellK07}
use abstract interpretation in combination with Gr\"obner bases computations for
computing polynomial invariants of bounded
degree. In~\cite{CacheraJJK12}, the abstract interpretation approach from~\cite{ipl/Muller-OlmS04} and the
constraint-based approach from~\cite{SankaranarayananSM04} is
combined, yielding a 
procedure for computing invariants of bounded degree without resorting to
Gr\"obner bases.

The techniques in~\cite{FarzanK15,KincaidBBR17,KincaidCBR18,KincaidBCR19}
approximate an arbitrary loop by a single-path loop and then apply recurrence
solving to infer nonlinear invariants. They include guards in loops and
conditionals in their reasoning, and are also able to infer
inequalities as loop invariant. 
A data-driven approach to invariant generation is given
in~\cite{esop/0001GHALN13} using the guess-and-check methodology. Linear
algebra is used to guess candidate invariants from data generated by concrete
program executions where an upper bound on the polynomial degree of the
candidate is user-given. An SMT solver is then used to validate the candidates
with respect to the properties of loop invariants. If this is not the case, then
the candidate is refined based on the output of the SMT solver.

Our work for invariant generation does neither use abstract interpretation nor
constraint solving, and does not fix an a priori bound on the degree of the
polynomial invariants to be synthesised. Instead, we 
restrict the class of loops our work can handle to non-deterministic loops
whose loop updates yield special classes of algebraic recurrences in
the loop counter, and hence we cannot handle loops with arbitrary
nestedness as in~\cite{KincaidCBR18}.
We rely on results of~\cite{Rodriguez-CarbonellK07} proving that the set of all polynomial equality
invariants for a given (non-deterministic) loop forms a polynomial
ideal. In~\cite{kovacs08}, we use the ideal-theoretic
result of~\cite{Rodriguez-CarbonellK07} and compute all polynomial
invariants of the class of non-deterministic 
loops that can be modelled by 
C-finite recurrence equations. Our results can further be extended to
more complex recurrences equations by allowing restricted
multiplications, and hence restricted classes of linear recurrences
with polynomial coefficients, among loop variables - as detailed
in~\cite{HumenbergerJK17,vmcai18}.

\paragraph{Algebra-based synthesis of loop invariants.}
We now summarise our algebra-based algorithm for synthesising
polynomial loop invariants. To this end, we define our task of loop invariant synthesis as follows: \\

\begin{mdframed}[frametitle=Loop Invariant Synthesis, 
       frametitlefont=\small\sffamily\MakeUppercase,
       innertopmargin=1pt,
       innerrightmargin=7pt,
       innerleftmargin=7pt,
       frametitleaboveskip=7pt,
       innerbottommargin=6pt
       ]
       \begin{itemize}
         \item[$\bullet$] \reserved{Given:} A non-deterministic single-path loop $\mathcal{L}$ with program
 variables $\vec{x}$ such that
each variable from $\vec{x}$ induces a C-finite number
sequence in $\mathcal{L}$; 
\item[$\bullet$]  \reserved{Generate:} A polynomial ideal $I$ of all polynomials $p(\vec{x})$
  such that  ${p(\vec{x})=0}$ is a loop invariant of $\mathcal{L}$. 
  \end{itemize}
 \end{mdframed}

The main steps of our algorithm for loop invariant synthesis are as follows:
\begin{enumerate}
    \item The non-deterministic single-path loop $\mathcal{L}$ is
      transformed into the regular expression $\pp^*$, where $\pp$ is
      the block of assignments from the loop body of $\mathcal{L}$ and $\pp^*$
    denotes an arbitrary number of executions of $\pp$. 
    \item We extract a system of C-finite recurrence equations for $\pp^*$, by
      describing 
    the C-finite number sequences for each program variable $x_i\in\vec{x}$ of $\mathcal{L}$
    via a C-finite recurrence equation. To this end, we write $x_i(n)$ to denote\label{page:notation}
    the value of the program variable $x_i\in\vec{x}$ at an arbitrary loop iteration $n\geq 0$ as well as to
    refer to the number sequence $x_i(n)$ induced by the values of $x_i$ at
    arbitrary loop iterations $n\geq 0$.

  \item We solve the 
resulting C-finite recurrences of  $\pp^*$, yielding a functional
representation of values of $x_i(n)$ depending only on $n$ and some
    initial values. 
    \item As a result, we derive closed forms 
      ${x_i(n) = f_i(n)}$, where $f_i$ are linear combinations of   polynomial
    and exponential expressions  in $n$.  We also compute algebraic
    relations $a_i(n)$ as valid polynomial relations among exponential
    expressions in $n$.
  \item A polynomial ideal $I$ of all polynomials $p(\vec{x})$
  such that  ${p(\vec{x})=0}$ is a loop invariant of $\mathcal{L}$ is
  then computed by using Gr\"obner basis computation to eliminate $n$
  from the ideal generated by $\id{x_i-f_i(n), a_i(n)}$. The ideal $I$
  is called the {\it polynomial invariant ideal} of $\mathcal{L}$. 
\end{enumerate}

\begin{example}[Loop invariant synthesis]
  We illustrate our algorithm for loop invariant
  synthesis on the loop of Figure~\ref{fig:motivating:a}. The loop
  guard of Figure~\ref{fig:motivating:a} is ignored. Using matrix
  notation, the block $\pp$ of loop body assignments induces the following coupled system
  of C-finite recurrence equations for $\pp^*$, with $n\geq 0$: 
  \begin{equation*}
    \begin{pmatrix}
      x(n+1) \\
      z(n+1) \\
      y(n+1)
    \end{pmatrix}=
    \begin{pmatrix}
      2 & 0 & 1 \\
      0 & 1 & 0 \\
      0 & 0 & 1
    \end{pmatrix}
    \begin{pmatrix}
      x(n) \\
      z(n) \\
      y(n)
    \end{pmatrix}
    +
    \begin{pmatrix}
      1 \\
      2 \\
      1
    \end{pmatrix}
  \end{equation*}
  The closed form solutions of the above recurrence system are given by
  \begin{equation*}
      \left\{
      \begin{array}{lcl}
          x(n) &=& x(0)+n^2\\
        z(n) &=& z(0)+2n\\
          y(n) &=& y(0)+n\\
      \end{array}\right.
  \end{equation*}
      with $x(0)=0$, $y(0)=0$ and $z(0)=0$ from the initial value assignments
      of Figure~\ref{fig:motivating:a}. By eliminating $n$ from
      $\id{x-n^2, z-2n,y-n}$, we derive the polynomial invariant ideal $I=\id{x-y^2, z-2y}$
  of $\pi^*$, yielding the polynomial loop invariant $x=y^2 \wedge  z=2y$.
  \end{example}

\paragraph{Automation and Implementation.}
Our algorithm for loop invariant synthesis is fully automated within
the open-source Julia package \aligator{}, which is available at: 
    \begin{center}
    \url{https://github.com/ahumenberger/Aligator.jl}.
    \end{center}
For experimental summary and comparisons with other tools, in
particular with~\cite{KincaidCBR18}, we refer to~\cite{cicm18,ThesisHumenberger20}.

\section{Algebra-based Synthesis of Loops}\label{sec:LoopSynt}

\paragraph{Overview of state-of-the-art.}
The classical setting of program synthesis has
been to synthesise programs from proofs of logical specifications that relate
the inputs and the outputs of the
program~\cite{MannaW80}. Thanks to recent successful trends in formal
verification based on automated reasoning~\cite{Z3,kovacsCAV13}, this
traditional view of program synthesis has been refined to the setting of
syntax-guided synthesis (SyGuS)~\cite{Alur15}.  In addition to logical
specifications, SyGuS approaches consider further constraints on the program
template to be synthesised, limiting thus the search space of possible
solutions. A wide range of efficient applications of SyGuS have so far 
emerged, for example programming by examples~\cite{GulwaniIJCAR16},
component-based synthesis~\cite{DBLP:conf/icse/JhaGST10} with learning~\cite{DilligPLDI18}
and sketching~\cite{SolarICML19}.

Most synthesis approaches 
exploit counterexample-guided
synthesis~\cite{Alur15,Solar09,DilligPLDI18,SolarICML19} within the
SyGuS framework. These
methods take  
input-output examples satisfying a given property and synthesise a
candidate program 
that is consistent with the given inputs. Correctness
of the candidate program  with respect to the given property is then
checked using formal verification, in particular using SMT-based
reasoning. Whenever verification fails, a counterexample violating the
given property is generated
as an additional input and a new candidate program is generated. Our
work does not use an iterative refinement of the input-output
values satisfying a given property $p(\vec{x})=0$. Rather, we consider a
precise 
characterisation of 
the solution space of loops with invariant
$p(\vec{x})=0$ to describe 
all, potentially infinite input-output values of interest.
Similarly to sketches~\cite{Solar09,SolarICML19}, we consider loop templates restricting the
search for solutions to synthesis. Yet, our templates support non-linear
arithmetic, which is not yet the case
in~\cite{SolarICML19,DilligPLDI18}.

The programming by example approach of~\cite{GulwaniPOPL11} learns programs from
input-output examples and relies on lightweight interaction to refine the
specification of programs to be synthesised. The approach has further been
extended in~\cite{GulwaniICLR18} with machine learning, allowing to learn
programs from just one (or even none) input-output example by using a simple
supervised learning setup. Program synthesis from input-output examples is shown
to be successful for recursive programs~\cite{GulwaniCAV13}, yet synthesising
loops and handling non-linear arithmetic is not yet supported by this
line of research. Our work precisely
characterises the solution space of all loops to be synthesised by a system of
algebraic recurrences and does not use statistical models supporting
machine learning.

To the best of our knowledge, existing synthesis approaches are restricted to linear
invariants, see e.g.~\cite{GulwaniPOPL10}, whereas our work supports
loop synthesis from non-linear polynomial properties. We note that many
interesting program
properties can be best expressed using non-linear arithmetic, for
example programs implementing powers (see
e.g. Figure~\ref{fig:motivating}), square roots and/or Euclidean
divison require non-linear invariants.

\paragraph{Algebra-based synthesis of loops.} 
Our work in~\cite{Humenberger20} addresses the challenging task of
loop synthesis, by 
relying on algebraic recurrence equations and constraint solving over
polynomials. Following the SyGuS setting, we consider additional
requirements on the loop to be synthesised and define the task of loop synthesis as follows: \\

\begin{mdframed}[frametitle=Loop Synthesis, 
       frametitlefont=\small\sffamily\MakeUppercase,
       innertopmargin=1pt,
       innerrightmargin=7pt,
       innerleftmargin=7pt,
       frametitleaboveskip=7pt,
       innerbottommargin=6pt
       ]
       \begin{itemize}
\item[$\bullet$] \reserved{Given:} A polynomial ideal $I$ containing
  polynomials $p(\vec{x})$ over a set $\vec{x}$ of 
  variables; 
  \item[$\bullet$] \reserved{Generate:} A loop $\mathcal{L}$ with program
  variables $\vec{x}$ such that
  \begin{itemize}
    \item[(i)] ${p(\vec{x})=0}$ is an invariant of $\mathcal{L}$ for every $p\in I$, and
    \item[(ii)] each variable from $\vec{x}$ in $\mathcal{L}$ induces a C-finite number
      sequence. 
    \end{itemize}
    \end{itemize}
  \end{mdframed}

 The main steps of our loop synthesis algorithm are summarised below.

 \begin{enumerate}
   
 \item We take a basis $B$ of the polynomial invariant ideal $I$ as our input.
   
    \item We fix a non-deterministic loop template $\mathcal{T}$
      whose loop updates define a C-finite recurrence system template 
      $\mathcal{S}$, over variables
      $\vec{x}$ and of size $s$. If not specified, the size $s$ of $\mathcal{S}$ is considered to
      be the number of variables in $\vec{x}$.
      
    \item We construct a polynomial constraint problem (PCP) which can be divided into
      two clause sets $C_1$ and $C_2$.
      The first set $C_1$ describes the closed
    form solutions of the C-finite recurrence system $\mathcal{S}$. To this end, we 
  exploit properties of C-finite recurrences  and define templates  
  for the closed forms of $\vec{x}$ by ensuring a one-to-one
  correspondence between the recurrence template $\mathcal{S}$ and
  the closed form templates of $\vec{x}$. Intuitively, the clause set $C_1$ mimics the procedure for computing the closed forms for the recurrence system $\mathcal{S}$.
    The second
    clause set $C_2$ of our PCP makes sure that, for every ${p\in B}$, $p(\vec{x})$ is an algebraic relation for the
    closed form templates of $\vec{x}$. Since $B$ is a basis of $I$ it follows that ${p(\vec{x})=0}$ for all $p\in I$.
The solution space of our PCP $C_1\wedge C_2$  
  captures thus the set of all C-finite recurrence systems of the
  form~$\mathcal{S}$ such that $p(\vec{x}(n))=0$ holds for
  all ${n\geq 0}$ and for all ${p\in I}$, where $\vec{x}(n)$ denotes the number sequences
  induced by the loop variables in $\vec{x}$ (as discussed
  on page~\pageref{page:notation}).
  
\item By solving our PCP, we derive C-finite recurrence systems of the
  form $\mathcal{S}$. These instances of $\mathcal{S}$ can however be
  considered as non-deterministic programs
      with simultaneous updates. Thus, any C-finite recurrence system
      solution of our PCP can directly be translated into a
non-deterministic loop  $\mathcal{L}$ with sequential updates, by introducing auxiliary
variables. Solving our PCP yields therefore a solution to our task of loop
synthesis. 
\end{enumerate}

In~\cite{Humenberger20}, 
 we prove that our approach to  loop
synthesis is both {sound} and {complete}.  By completeness we
mean, that 
if there is a loop $\mathcal{L}$ with at most $s$ variables satisfying the invariant ${p(\vec{x})=0}$ such that the loop
body meets the C-finite syntactic requirements of $\mathcal{S}$, then
this loop $\mathcal{L}$ is synthesised by our method.
As show-cased by  Figure~\ref{fig:Dafny}, given a  loop invariant ${p(\vec{x})=0}$, one can
synthesise a potentially infinite set of loops such that each loop (i) has ${p(\vec{x})=0}$
as its invariant and (ii) is ``better'' with respect to a
user-defined preference/measure. Our loop synthesis approach can thus
be used to synthesise loops with respect to some
pre-defined measure. 

\begin{example}[Loop invariant synthesis]
We illustrate our algorithm for loop synthesis on
Figure~\ref{fig:motivating:b}.
To this end, we are interested in synthesising loops from the non-linear
polynomial relation ${x=y^2}$. The invariant we consider is
  therefore ${p(x,y)=x-y^2=0}$.

  We start by (initially) setting $s=2$ and defining a loop template $\mathcal{T}$ of the form
  \begin{equation}
    \label{motivating:template}
    \begin{tabular}{l}
    $(x, y) \gets (a_1, a_2)$\\
    $\WHILE~true~\DO$\\
      \quad$x \gets b_{11}x + b_{12}y + b_{13}$\\
    \quad$y \gets b_{21}x + b_{22}y + b_{23}$\\
    $\END$
    \end{tabular}
  \end{equation}
  where the $a_i$ and $b_{ij}$ are rational-valued symbolic constants. 
  By denoting with $n\geq 0$ the loop counter,
  the  loop body of~\eqref{motivating:template} can then
  be modeled by the following C-finite recurrence system:
  \begin{equation}
    \label{motivating:rec-template}
    \begin{pmatrix}
      x(n+1) \\
      y(n+1)
    \end{pmatrix}=
    \begin{pmatrix}
      b_{11}' & b_{12}' \\
      b_{21}' & b_{22}'
    \end{pmatrix}
    \begin{pmatrix}
      x(n) \\
      y(n)
    \end{pmatrix}
    +
    \begin{pmatrix}
      b_{13}' \\
      b_{23}'
    \end{pmatrix},
  \end{equation}
  where $x(n)$ and $y(n)$ represent the values of variables $x$ and $y$ at
  iteration $n$ (as discussed on page~\pageref{page:notation}), with ${x(0)=a_1}$ and ${y(0)=a_2}$. 
Note that the values of
  $b_{ij}$ and $b_{ij}'$ might differ as the sequential
  assignments of~\eqref{motivating:template} correspond to
  simultaneous assignments in the algebraic
  representation~\eqref{motivating:rec-template} of the loop.

  We next
  exploit properties of C-finite recurrences. For simplicity and
  w.l.o.g, we set
  up the following closed form templates  
  for  $x(n)$ and $y(n)$:
  \begin{equation}
    \label{eq:cf-template}
    \begin{pmatrix}
      x(n) \\ y(n)
    \end{pmatrix} =
    \begin{pmatrix}
      c_1 \\ c_2
    \end{pmatrix} \omega^n +
    \begin{pmatrix}
      d_1 \\ d_2
    \end{pmatrix} \omega^n n +
    \begin{pmatrix}
      e_1 \\ e_2
    \end{pmatrix} \omega^n n^2
  \end{equation}
where $c_i,d_i,e_i$ are   rational-valued symbolic constants and
$\omega$ are symbolic algebraic numbers. 
  We then generate the clause set $C_1$ that ensures that we have a one-to-one correspondence between the number sequences described by the recurrence equations and the closed forms.
  For making sure that the equation ${x-y^2=0}$ is indeed a polynomial invariant, we plug the closed form templates~\eqref{eq:cf-template} into the equation, and get
  \begin{equation}
    \label{eq:algrel}
    c_1\omega^n + d_1 \omega^n n + e_1\omega^n n^2 - (c_2\omega^n + d_2 \omega^n n + e_2\omega^n n^2)^2=0.
  \end{equation}
  The above equation~\eqref{eq:algrel} has to hold for all ${n\in\N}$ as
  ${x-y^2=0}$ should be a loop invariant. That is, we want to find
  $c_1,c_2,d_1,d_2,e_1,e_2$ and $\omega$ such that~\eqref{eq:algrel} holds for
  all ${n\in\N}$. The properties of C-finite number sequences allow us to reduce
  this $\exists\forall$ problem containing exponential expressions into a finite
  set of polynomials 
  \begin{equation*}
    \begin{aligned}
     C_2 = \{ &c_1 \omega - c_2^2 \omega^2=0,
      d_1 \omega - 2 c_2 d_2 \omega^2=0,\\
      &e_1 \omega - (2 c_2 e_2 - d_2^2) \omega^2=0,
      2 d_2 e_2 \omega^2=0,
      e_2^2 \omega^2=0 \}
    \end{aligned}
  \end{equation*}

  In summary, we get a PCP consisting of clause sets $C_1$ and $C_2$ containing
  {27} polynomial constraints over the unknowns $a_{i}, b'_{ij}, c_i, d_i, e_i,
  \omega$ from~\eqref{motivating:template}-\eqref{eq:cf-template}. The solution
  space of our PCP captures the set of all C-finite recurrence systems of the
  form~\eqref{motivating:rec-template} such that ${x(n)-2y(n)^2=0}$ holds for
  all ${n\geq0}$.
  That is, any solution of our PCP yields a loop with
  an invariant $x=y^2$. 
  
  Figures~\ref{fig:Dafny}(b)-(c) illustrate two solutions of the PCP problem of
  our example: each program of Figure~\ref{fig:Dafny}(b)-(c) is an instance
  of~\eqref{motivating:template}, has ${x-2y^2=0}$ as its invariant and can be
  synthesised using our work. The loop of Figure~\ref{fig:Dafny}(b),
  and thus of Figure~\ref{fig:motivating:b}, is synthesised by
  considering the size $s$ of~\eqref{motivating:template} to be $2$,
  whereas Figure~\ref{fig:Dafny}(c) is computed by increasing the size
  $s$ of~\eqref{motivating:template} to 3.

\end{example}

\paragraph{Automation and Implementation.}  We implemented our approach to loop synthesis in the new open-source Julia package
    \absynth{}, available at
    \begin{center}
        \url{https://github.com/ahumenberger/Absynth.jl}.
    \end{center}
Our experiments using academic benchmarks on loop analysis as
    well as on generating number sequences in algorithmic
    combinatorics are available in~\cite{Humenberger20,ThesisHumenberger20}.

\section{Conclusions}
We overviewed algebra-based algorithms for loop invariant synthesis and loop synthesis.
The key ingredient of our work comes by modeling loops as algebraic
recurrences, in particular by C-finite recurrences. To this end, we
consider non-deterministic loops whose loop updates induce C-finite
number sequences among loop variables.  In the case of
loop invariant synthesis, our work generates the polynomial ideal of all
polynomial invariants of such loops by using symbolic summation in
combination with properties of polynomial ideals. Extending this
approach to (multi-path) loops inducing more complex recurrence
equations supporting for example arbitrary multiplications 
 among (some of the)
variables is an interesting line for future work.
When synthesising loops from polynomial invariants,
we use symbolic summation to generate polynomial constraints whose
solutions yield loops that exhibit the given invariant. Solving our constraint
system requires satisfiability solving in non-linear arithmetic,
opening up new directions for SMT-based reasoning with polynomial
constraints. For example, we believe searching for solutions over
finite domains would improve the scalability of our loop synthesis
method. Extending our loop synthesis task to generate
loops that are optimal with respect to a user-specified measure is
another challenge to further investigate. To this end, understanding
and efficiently  encoding the best optimisation measures into our approach
is an interesting line for future work.
\bigskip

\noindent{\bf{Acknowledgments.}}
  We thank Maximillian Jaroschek (TU Wien) for joint work allowing to
  extend our invariant generation approaches to more complex loops and
  number sequences.  
  We also thank 
Sumit Gulwani (Microsoft) and Manuel Kauers (JKU Linz) for valuable discussions on ideas leading
to our loop synthesis framework. 
Practical aspects of our loop synthesis approach and using 
loop synthesis for strength reduction involve joint
work with Nikolaj Bj\o{}rner (Microsoft). 

We acknowledge funding from the ERC Starting Grant 2014
SYMCAR 639270, the ERC Proof of Concept Grant SYMELS 842066, the Wallenberg Academy
Fellowship TheProSE, and the Austrian FWF research project W1255-N23.
\bibliographystyle{plain}
\bibliography{refs}

\end{document}